\documentclass[showpacs,amsmath,amssymb]{revtex4}
\usepackage{amsmath}
\usepackage{amssymb}
\usepackage{graphicx}
\begin{document}
\title{Giant dipole resonance with exact treatment of thermal fluctuations}
\author{P. Arumugam$^1$} \email{aru@iopb.res.in}
\altaffiliation[Present address: ]{Institute of Physics,
Sachivalaya Marg, Bhubaneswar - 751 005, India.}
\author{G. Shanmugam$^1$}
\author{S.K. Patra$^2$}
\affiliation{$^1$ Department of Physics, Manonmaniam Sundaranar
University, Tirunelveli - 627 012, India.  \\ $^2$ Institute of
Physics, Sachivalaya Marg, Bhubaneswar - 751 005, India.}
\date{January 12, 2004}

\begin{abstract}
The shape fluctuations due to thermal effects in the giant dipole
resonance (GDR) observables are calculated using the exact free
energies evaluated at fixed spin and temperature. The results
obtained are compared with Landau theory calculations done by
parameterizing the free energy. The Landau theory is found to be
insufficient when the shell effects are dominating.
\end{abstract}

 \maketitle

The study of structural transitions as a function of both angular
momentum and temperature has been one among the fascinating
aspects of highly excited nuclei in recent years
\cite{Heck03,Came03,Brac03,Kusn03}.  The Giant Dipole Resonance
(GDR) studies have been proved to be a powerful tool to study such
hot and rotating nuclei \cite{GDRREV} and recently the domain of
GDR spreads rapidly over different areas of theoretical and
experimental interest \cite{Maj95,Came01,Tson00,Dios01}.  The GDR
observations provide us information about the geometry as well as
the dynamics of nuclei even at extreme limits of temperature
($T$), spin ($I$) and isospin ($\tau$). In the past most of the
GDR measurements in hot nuclei were made at moderate and high $T$
\cite{GDRREV,Maj95,Came01,Tson00,Dios01,Gund90,Baum98}. Several
experiments have been carried out recently to study the GDR states
at low temperatures \cite{Heck03,Came03,Brac03}. Hence the
theories which were successful in the high $T$ regime should be
now scrutinized with the low temperature observations as well as
with theories incorporating properly the microscopic effects (such
as shell effects) which are dominant at low temperatures.  While
dealing with the thermal shape fluctuations, free energy
parameterizations such as Landau theory \cite{ALHA,ALHAE,SELVP}
are usually employed to do timesaving calculations.  In this work
we survey the applicability of Landau theory by demanding
consistency with exact calculations done without any parameter
fitting.

The theoretical approach we follow is of three fold with models
for 1) shape calculations, 2) relating the shapes to GDR
observables and 3) considering the shape fluctuations due to
thermal effects. For shape calculations we follow the
Nilsson-Strutinsky (NS) method extended to high spin and
temperature \cite{SELVP,RAMS,ARU2}. The total free energy
($F_\mathrm{TOT}$) at fixed deformation is calculated using the
expression
\begin{equation}\label{FTOT}
F_\mathrm{TOT}=E_\mathrm{LDM}+\sum_{p,n}\delta F +\frac12 \omega
(I_\mathrm{Classical}+\sum_{p,n}\delta I)\;.
\end{equation}
The liquid-drop energy ($E_\mathrm{LDM}$) is calculated by summing
up the Coulomb and surface energies \cite{RAMS} corresponding to a
triaxially deformed shape defined by the deformation parameters
$\beta$ and $\gamma$.  The classical part of spin
($I_\mathrm{Classical}$) is obtained from the rigid-body moment of
inertia with surface diffuseness correction \cite{RAMS}.  The
shell correction ($\delta F$) is the difference between the
deformation energies evaluated with a discrete single-particle
spectrum and by smoothing (averaging) that spectrum ($\delta
F=F-\widetilde{F}$). Similarly the shell correction corresponding
to the spin is given by $\delta I=I-\widetilde{I}$.  To calculate
the shell corrections for energy and spin, we use the triaxially
deformed Nilsson model together with the Strutinsky's
prescription. The single-particle energies ($e_i$) and spin
projections ($m_i$) are obtained by diagonalizing the triaxial
Nilsson Hamiltonian in cylindrical representation upto first
twelve major shells.  At finite temperatures the free energy is
given by
\begin{equation}
\label{FT}F=\sum_{i=1}^\infty e_in_i-T\sum_{i=1}^\infty s_i\,,
\end{equation}
where $s_i$ are the single-particle entropy and $n_i$ are the
occupation numbers which follow Fermi-Dirac distribution given by
\begin{equation}
\label{NI}n_i=\frac 1{1+\exp \left( \frac{e_i-\lambda }T\right)
}\;.
\end{equation}
The chemical potential $\lambda$ is obtained using the constraint
$\sum_{i=1}^\infty n_i = N$, where $N$ is the total number of
particles.  The total entropy $S=\sum_{i=1}^\infty s_i$ can be
represented in terms of occupation numbers as
\begin{equation}
S=-\sum_{i=1}^\infty\left[n_i\ln n_i-(1-n_i)\ln (1-n_i)\right] \;.
\end{equation}

The shell correction energy at finite temperature is calculated
using the expression \cite{ARU2,BQ81}
\begin{equation}
\label{FTT2}\widetilde F=\sum_i e_i\widetilde
n_i-T\sum_i\widetilde s_i \;,
\end{equation}
where $\widetilde n_i=\int_{-\infty }^\infty \widetilde
f(x)\;n_i(x)\;\d x \;,$ and $\widetilde s_i=\int_{-\infty }^\infty
\widetilde f(x)\;s_i(x)\;\d x$. The integrals appearing in
Eq.(\ref{FTT2}) are evaluated numerically using the Hermite-Gauss
quadrature.  Apart from the numerical evaluation of integrals,
this method gives exact temperature dependent shell corrections.
For the spin distribution, the Strutinsky smoothed spin can be
derived in a similar way leading to the expression
$\widetilde{I}=\sum_{i=1}^\infty m_i \widetilde{n}_i$ and hence
the shell correction for spin is
\begin{equation}
\delta I=\sum_{i=1}^\infty m_i n_i - \sum_{i=1}^\infty m_i
\widetilde{n}_i \;.
\end{equation}
The most probable shapes are obtained by minimizing the total free
energy (\ref{FTOT}).

The nuclear shapes are related to the GDR observables using a
model \cite{THIAG} comprising an anisotropic harmonic oscillator
potential with separable dipole-dipole interaction.  In this
formalism the GDR frequencies in laboratory frame are obtained as
\begin{equation}
\widetilde{\omega}_z=(1+\eta )^{1/2}\omega _z \;,
\end{equation}
\begin{eqnarray}
\nonumber
\widetilde{\omega}_2\mp\Omega&=&\left\{(1+\eta )\frac{\omega _y^2+\omega _x^2}%
2\;+\Omega^2 + \frac12\left[ (1+\eta )^2(\omega _y^2-\omega
_x^2)^2 \right. \right. \\ && \left. \left. +8\Omega ^2 (1+\eta
)(\omega _y^2+\omega _x^2)\right] ^{\frac 12}\right\} ^{\frac
12}\mp \Omega \;,
\end{eqnarray}
\begin{eqnarray}
\nonumber
\widetilde{\omega}_3\mp\Omega&=&\left\{ (1+\eta )\frac{\omega _y^2+\omega _x^2}%
2\;+\Omega ^2 -\frac 12\left[ (1+\eta )^2(\omega _y^2-\omega
_x^2)^2 \right. \right. \\&& \left. \left. +8\Omega ^2(1+\eta
)(\omega _y^2+\omega _x^2)\right] ^{\frac 12}\right\} ^{\frac
12}\mp\Omega \;,
\end{eqnarray}
where $\Omega$ is the cranking frequency, $\omega_x,\ \omega_y,\
\omega_z$ are the oscillator frequencies derived from the
deformation of the nucleus and $\eta$ is a parameter that
characterizes the isovector component of the neutron and proton
average field.  The GDR cross sections are constructed as a sum of
Lorentzians given by
\begin{equation}
\sigma (E_\gamma)=\sum_{i}
\frac{\sigma_{mi}}{1+\left(E_\gamma^2-E_{mi}^2\right)
^2/E_\gamma^2\Gamma_{i}^2}
\end{equation}
where Lorentz parameters $E_m$, $\sigma _m$ and $\Gamma$ are the
resonance energy, peak cross-section and full width at half
maximum respectively. Here $i$ represents the number of components
of the GDR and is determined from the shape of the nucleus
\cite{THIAG,HILT}.  It is to be noted that these Lorentz lines are
non-interfering, but $\Gamma_i$ is assumed to depend on energy.
The energy dependence of the GDR width can be approximated by
\cite{CARL}
\begin{equation}\label{PowLaw}
\Gamma_i \approx 0.026E_i ^{1.9} \;.
\end{equation}
The peak cross section $\sigma_m$ is given by
\begin{equation}
 \sigma_m=60\frac2\pi \frac {NZ}{A} \frac{1}{\Gamma} \;
 0.86(1+\alpha) \;.
\end{equation}
The parameter $\alpha$ which takes care of the sum rule is fixed
at 0.3 for all the nuclei considered in this work.  This parameter
has more effect on the peak cross section.  In most of the cases
we normalize the peak with the experimental data and hence the
choice of $\alpha$ has less effect on the results.  The other
parameter $\eta$ varies with nucleus so that the ground state GDR
centroid energy is reproduced. The choice for $^{90}$Zr, $^{92}$Mo
is $\eta=2.6$ and for $^{208}$Pb it is $\eta=3.4$.  For
calculating the GDR width, only the power law (\ref{PowLaw}) is
used in this work and no ground state width is assumed.

The relation between nuclear shape and GDR cross section is not
straightforward especially in hot nuclei where large-amplitude
thermal fluctuations of the nuclear shape play an important role
\cite{ALHAT}.  When the nucleus is observed at finite excitation
energy, the effective GDR cross-sections carry information on the
relative time scales for shape rearrangements. Hence in the case
of hot nuclei, for a meaningful comparison of experimental and
theoretical values, the thermal shape fluctuations should be taken
care properly.  In the case of hot and rotating nuclei there can
be fluctuations in the orientation of the nuclear symmetry axis
with respect to the rotation axis. The general expression for the
expectation value of an observable $\mathcal{O}$ incorporating
both thermal and orientation fluctuations is given by \cite{ALHA,ALHAO}
\begin{equation} \label{EqOFTF}
\langle \mathcal{O} \rangle_{\beta,\gamma,\Omega} = \frac{\int
\mathcal{D}[\alpha]e^{-F(T,I;\beta,\gamma,\Omega)/T}
(\hat{\omega}\cdot\mathcal{I}\cdot\hat{\omega})^{-3/2}
\mathcal{O}} {\int
\mathcal{D}[\alpha]e^{-F(T,I;\beta,\gamma,\Omega)/T}
(\hat{\omega}\cdot\mathcal{I}\cdot\hat{\omega})^{-3/2}} \;,
\end{equation}
where $\Omega=(\phi,\theta,\psi)$ are the Euler angles specifying
the intrinsic orientation of the system,
$\hat{\omega}\cdot\mathcal{I}\cdot\hat{\omega}=I_{x^\prime
x^\prime}\cos^2\phi\ \sin^2\theta + I_{y^\prime
y^\prime}\sin^2\phi\ \sin^2\theta + I_{z^\prime z^\prime}
\cos^2\theta$ is the moment of inertia about the rotation axis
$\hat{\omega}$ given in terms of the principal moments of inertia
$I_{x^\prime x^\prime},\ I_{y^\prime y^\prime},\ I_{y^\prime
x^\prime}$, and the volume element $\mathcal{D}[\alpha] = \beta^4
|\sin 3\gamma| \, d\beta \, d\gamma \, \sin\theta \, d\theta \,
d\phi$.

The study of thermal fluctuations by numerical evaluation of Eq.
(\ref{EqOFTF}) in general requires an exploration of five
dimensional space spanned by the deformation and orientation
degrees of freedom, in which a large number of points are required
in order to assure sufficient accuracy (especially at finite
angular momentum).  Hence certain parameterizations were developed
\cite{ALHAE,Orm97} to represent the free energy using functions
that mimic the behaviour of the NS calculation as closely as
possible. One such parameterization is the Landau theory of phase
transitions, developed by Alhassid {\it et al} \cite{ALHAE}.  Here
the free energy is expanded in terms of certain temperature
dependent constants which are to be extracted by fitting with the
free energy calculations at fixed temperatures from the NS method.
Moreover, once the fits involving free energy and moment of
inertia are made for the non-rotating case, the calculations can
be extended to higher spins using the relation \cite{ALHA}
\begin{equation} \label{EqFTI}
F(T,I;\beta,\gamma,\Omega)=F(T,\omega=0;\beta,\gamma)+\frac{(I+1/2)^2}
{2\ \hat{\omega}\cdot\mathcal{I}\cdot\hat{\omega}} \;.
\end{equation}
Hence this theory offers an economic parameterization to study the
hot rotating nuclei.  We have employed Landau theory in its
extended form as given in Refs. \cite{ALHAE,SELVP}.

With recent computing facilities, it is possible to perform the
thermal fluctuation calculations exactly by computing the
integrations in Eq. (\ref{EqOFTF}) numerically with the free
energies and the observables being calculated exactly at the
integration (mesh) points. In this way the calculations can be
done more accurately without using any parameterization and
consequent fitting.  In this work we have performed such
calculations, however, neglecting the orientation fluctuations.
This enables us to perform the integration in the deformation
space only which at present is two dimensional having the
deformation parameters $\beta$ and $\gamma$.  A 32 point Gaussian
quadrature has been used to evaluate the integrals numerically.
While performing fluctuation calculations in this way, the free
energy, GDR cross section and width at any given spin is obtained
by tuning the cranking frequency to get the desired spin. It has
to be noted that the NS free energy calculations at mesh points
($32\times 32$) have to be done only once for a fixed spin and
temperature while calculating the GDR cross sections at fixed spin
and temperature.

\begin{figure*}
\centering
\includegraphics[width=0.5\columnwidth, clip=true]{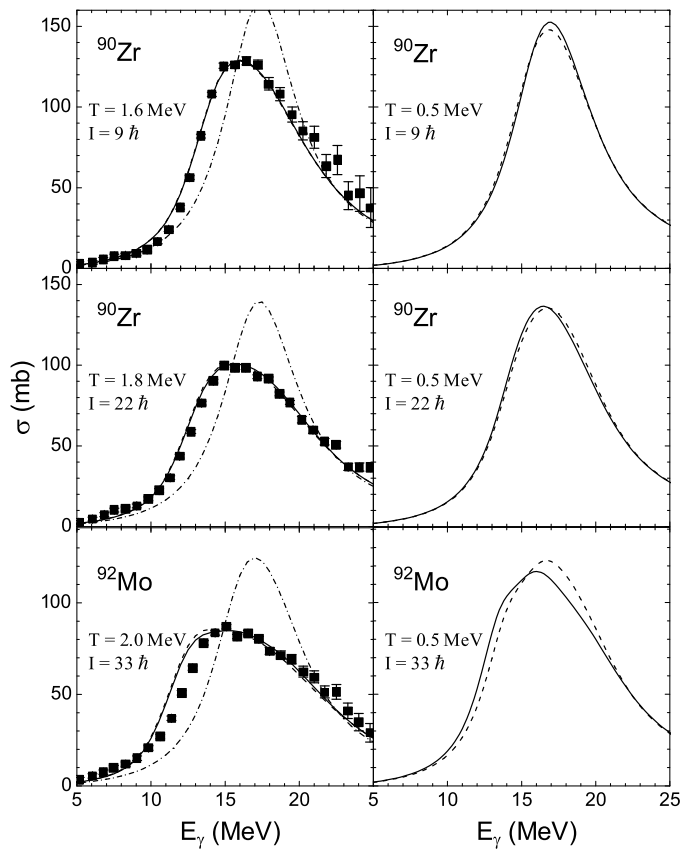}
\caption{GDR cross sections for the nuclei $^{90}$Zr and
$^{92}$Mo. \textit{Left:} Experimental data represented by solid
squares are taken from ref. \protect\cite{Gund90}. The solid lines
represent calculations with orientation and thermal fluctuations,
dashed lines correspond to thermal fluctuations alone and the
dash-dotted line correspond to most probable shapes.
\textit{Right:} Results obtained at $T=0.5$ MeV with Landau theory
(dashed lines) and exact method (solid line).}
\end{figure*}

\begin{figure*}
\centering
\includegraphics[width=0.4\columnwidth, clip=true]{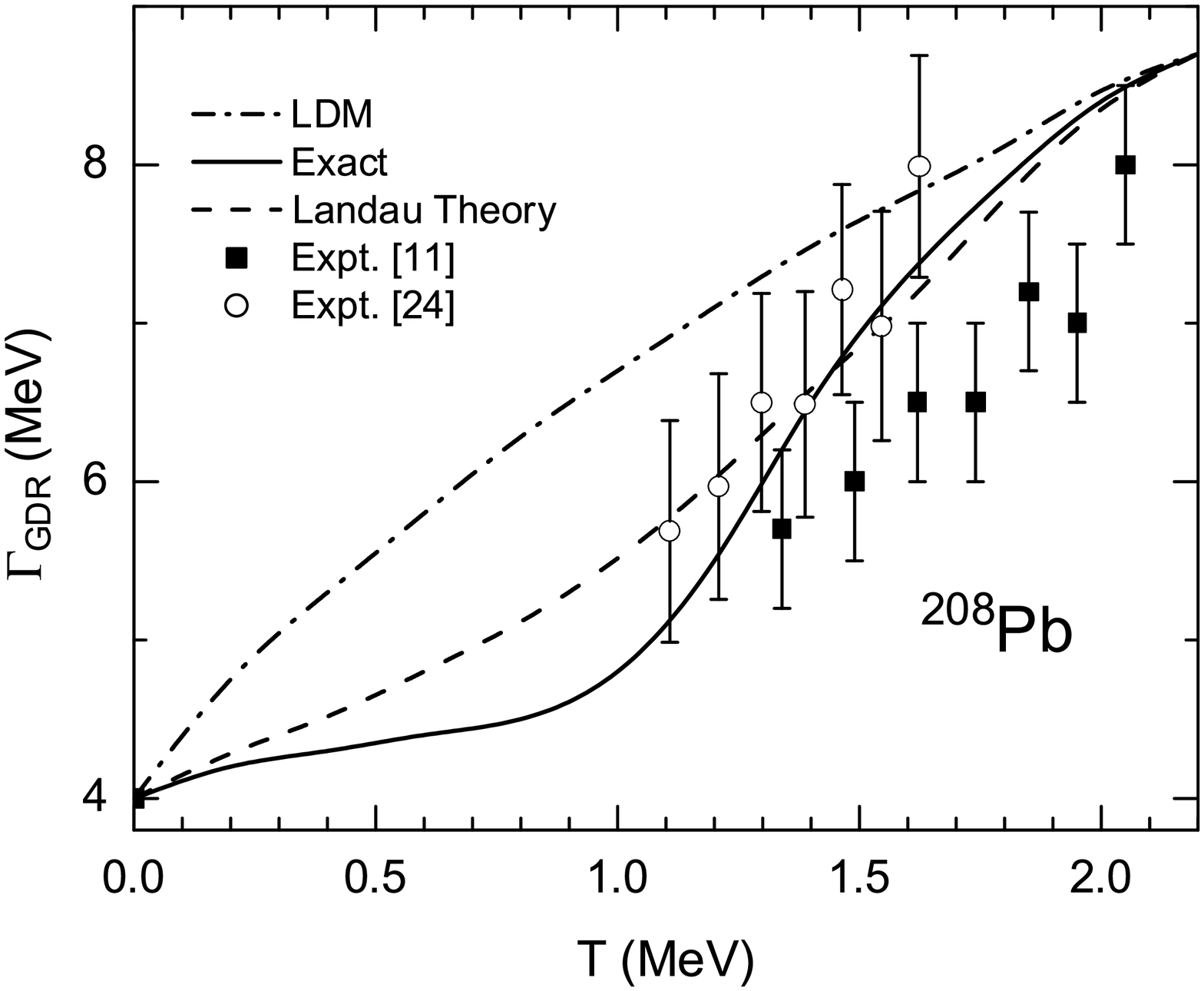}
\caption{GDR width in $^{208}$Pb. The results obtained using
liquid drop model (dash-dotted line), Landau theory (dashed line)
and the exact calculations (solid line) are compared. Experimental
data represented by solid squares are taken from ref.
\cite{Baum98} and the revised data \cite{Kusn98} are represented
by open circles.}
\end{figure*}

Now we compare our calculated results obtained by using the
extended Landau theory and the exact method.  In Fig.\ 1 as the
representative cases we show for the hot rotating $^{90}$Zr and
$^{92}$Mo nuclei, the results of our GDR cross section
calculations along with the experimental results \cite{Gund90}. At
experimentally observed temperatures, the most probable shapes are
found to be $\beta$=0.0, 0.09 and 0.1 for the spins 9$\hbar$,
22$\hbar$ and 33$\hbar$ respectively, with $\gamma=-180^o$ for all
the spins. The thermally averaged value of the deformation,
$\bar{\beta}$ is found to be 0.29, 0.33 and 0.41 for the spins
9$\hbar$, 22$\hbar$ and 33$\hbar$ respectively, and the averaged
$\gamma$ is found to be $\bar{\gamma}\approx148^o$. These values
correspond to Landau theory calculations and are consistent with
those reported earlier \cite{ALHAO}. As similar to previous
observations \cite{ALHAO} it is clear from Fig.\ 1 that thermal
fluctuations play crucial role whereas the orientation
fluctuations are negligible while calculating the GDR cross
sections.  Also this justifies the omission of orientation
fluctuations in our exact method. Importantly we find that the
results of the thermal fluctuation calculations using the
macroscopic extended Landau theory and the present exact approach,
are exactly similar (they exactly overlap in left panel of Fig.\
1) at the experimentally observed temperatures.  This equivalence
is in spite of the differences in the free energies, from NS
calculations and Landau theory, at larger deformations. The
thermal fluctuations are strong enough to make the GDR cross
sections insensitive to minor changes in the free energy surfaces.
We examined the situation at lower temperatures and the cross
sections calculated using two methods at $T=0.5$ MeV are shown in
the right panel of Fig.\ 1.  To enable better comparison we show
results without orientation fluctuations. It is evident that the
Landau theory results deviate at $I=33\hbar$. This can be ascribed
to the spin driven shell effects.  The rapid change in the
single-particle level structure with spin is well known and this
may lead to sharp changes in the shell corrections. The Landau
theory could not account for this as the free energies at higher
spins are obtained from the free energies and moment of inertia at
$I=0$ using Eq.\ (\ref{EqFTI}).

In Fig.\ 2 we present the calculated GDR widths of $^{208}$Pb at
$I=0\hbar$ along with experimental results.  The calculations are
performed with 1) the liquid-drop model (LDM) free energies and
Landau theory, 2) NS free energies and Landau theory and 3) NS
free energies with exact treatment of fluctuations.  The strong
shell corrections for spherical shape results in large difference
in the deformation energies between spherical and deformed
configurations.  This leads to attenuation of thermal fluctuations
at lower temperature and hence the obtained widths are much lower
when compared to liquid-drop model results.  The magnitude of this
attenuation comes out to be different in methods 2 and 3.  In the
presence of strong shell effects at lower temperatures, the Landau
theory results deviate considerably from the results of exact
calculations. Above $T\sim 1.5$MeV the results from two methods
are the same. The deviations from the NS calculations in the free
energy obtained by Landau theory at larger deformations
\cite{Orm97} are not reflected in the GDR observables calculated
at higher temperatures due to the thermal fluctuations and the
weakening of shell effects.  The results of our calculations
suggest that the parameterization of Landau theory is good enough
to explain the GDR properties of hot rotating nuclei in the
absence of strong shell effects. The shell effects can be treated
in a better way using the parameterization suggested in Ref.
\cite{Orm97}. However, spin driven shell effects cannot be
explained in such a formalism also as the difference in energy due
to rotation comes just through the moment of inertia.

To summarize, in this work the thermal fluctuations are dealt in
an exact way without any parameter fitting.  We have carried out a
case study of the GDR properties in the nuclei $^{90}$Zr,
$^{92}$Mo and $^{208}$Pb and our results are well in conformity
with experimental results. Comparison of our present approach with
the thermal fluctuation model comprising Landau theory has been
brought out.  The Landau theory is found to be insufficient to
explain GDR properties in the presence of strong shell effects.

\begin{acknowledgments}
One of the authors (PA) acknowledge financial support from Council
of Scientific and Industrial Research, Govt. of India under the
scheme CSIR-SRF: 9/652(10)/2002-EMR-I.
\end{acknowledgments}


\begin{thebibliography}{99}

\bibitem{Heck03}P. Heckman \textit{et al}, Phys. Lett. \textbf{B555} (2003) 43.

\bibitem{Came03}F. Camera \textit{et al}, Phys. Lett. \textbf{B560} (2003) 155.

\bibitem{Brac03} A. Bracco, Acta Phys. Pol. \textbf{B34} (2003) 2163.

\bibitem{Kusn03}D. Kusnezov, and W.E. Ormand, Phys. Rev. Lett. \textbf{90} (2003)
042501.

\bibitem{GDRREV} K.A. Snover, Annu. Rev. Nucl. Part. Sci.
\textbf{36} (1986) 545; J.J. Gaardh\o je, Annu. Rev. Nucl. Part.
Sci. \textbf{42} (1992) 483.

\bibitem{Maj95} A. Maj \textit{et al}, Acta Phys. Pol. \textbf{B26} (1995) 417.

\bibitem{Came01} F. Camera \textit{et al}, Acta Phys. Pol. \textbf{B32} (2001) 807.

\bibitem{Tson00} N. Tsoneva, Ch. Stoyanov, Yu.P. Gangrsky, V.Yu. Ponomarev, N.P. Balabanov, and A.P. Tonchev,
Phys. Rev. \textbf{C 61} (2000) 044303.

\bibitem{Dios01}I. Di\'{o}szegi, I. Mazumdar, N.P. Shaw, and P. Paul,
Phys. Rev. \textbf{C 63} (2001) 047601.

\bibitem{Gund90} J.~Gundlach, K.A. Snover, J.A. Behr, C.A. Gossett, M.K. Habior, and K.T. Lesko,
Phys. Rev. Lett. {\bf 65} (1990) 2523.

\bibitem{Baum98}
T. Baumann \textit{et al}, Nucl. Phys. {\bf A635}, 428 (1998).

\bibitem{ALHA}  Y. Alhassid, Nucl. Phys. {\bf A649} (1999) 107c.

\bibitem{ALHAE}  Y. Alhassid and B. Bush, Nucl. Phys. {\bf A549} (1992) 12.

\bibitem{SELVP}  G.~Shanmugam and V.~Selvam, Phys. Rev. {\bf C 62}, (2000) 014302.

\bibitem{RAMS} G. Shanmugam, V. Ramasubramanian and P. Arumugam, Pramana - J. Phys. {\bf 53}
(1999) 457; G. Shanmugam, V. Ramasubramanian and
S.N.~Chintalapudi, Phys. Rev. {\bf C 63} (2001) 064311.

\bibitem{ARU2} G. Shanmugam and P. Arumugam, Pramana - J. Phys. {\bf 57}
(2001) 223.

\bibitem{BQ81}  M. Brack and P. Quentin, Nucl. Phys. {\bf A361} (1981) 35.

\bibitem{THIAG} G. Shanmugam and M. Thiagasundaram, Phys. Rev.
{\bf C 37} (1988) 853; \textit{ibid}, Phys. Rev. {\bf C 39} (1989)
1623.

\bibitem{HILT} R.R. Hilton, Z. Phys. {\bf A 309} (1983) 233.

\bibitem{CARL} P. Carlos, R. Bergere, H. Beil, A. Lepretre and
A.~Veyssiere, Nucl. Phys. {\bf A219}, 61 (1974).

\bibitem{ALHAT}  Y. Alhassid and B. Bush, Nucl. Phys. {\bf A509} (1990)
461.

\bibitem{ALHAO}  Y. Alhassid and B. Bush, Nucl. Phys. {\bf A531} (1991) 39.

\bibitem{Orm97} W.E. Ormand, P.G. Bortignon and R.A. Broglia, Nucl.
Phys. {\bf A614} (1997) 217.

\bibitem{Kusn98}
D. Kusnezov, Y. Alhassid and K.A. Snover, Phys. Rev. Lett.
\textbf{81} (1998) 542.

\end{thebibliography}
\end{document}